\documentstyle[twocolumn,eqsecnum,aps]{revtex}
\begin{document}
\draft
\preprint{HEP/123-qed}

\title{Stability Issues in Euclidean Quantum Gravity}

\author{G. Modanese \cite{byline}}

\address{I.N.F.N., Gruppo Collegato di Trento \\
I-38050 Povo (TN) - Italy \\ and \\
European Centre for Theoretical Studies in Nuclear Physics and 
Related Areas \\ Villa Tambosi, Strada delle Tabarelle 286 \\ I-38050 
Villazzano (TN) - Italy}

\maketitle

\begin{abstract}

It is known that the action of Euclidean Einstein gravity is not
bounded from below and that the metric of flat space does not correspond
to a minimum of the action.  Nevertheless, perturbation theory about flat
space works well. The deep dynamical reasons for this reside in the
non-perturbative behaviour of the system and have been clarified in part
by numerical simulations.  Several open issues remain. We treat in
particular those zero modes of the action for which $R(x)$ is not
identically zero, but the integral of $\sqrt{g(x)} R(x)$ vanishes. 

\end{abstract}

\pacs{PACS Numbers: 04.20.-q, 04.60-m}

\section{Introduction.}
\label{int}

The Euclidean (or ``imaginary time") formulation of a quantum field theory
offers several aesthetic and substantial advantages. The practical rules
for perturbative computations are simpler than in the Lorentzian case and
there is no distinction between upper and lower indices. The only relevant
Green function of the linearized equations is the Feynman propagator, and
there is no need of formal regularization through the $i \varepsilon$
term.  From the non-perturbative point of view, if the Euclidean action
$S$ is positive definite, then the functional integral is formally
convergent thanks to the exponential of $-S/\hbar$. 

When time-independent quantities are computed in the Euclidean theory, the
inverse analytical continuation to real time is not necessary.  A well
known example are the formulas for the static potential \cite{sym}. We
recall them in some detail in Section \ref{rec}. 

The physical correspondence between an Euclidean functional integral and a
statistical system at the temperature $\Theta=\hbar/k_B$ is immediate. We
can also easily visualize the dynamics of the system, after suitable
discretization, as a Montecarlo evolution: starting from a given field
configuration, a new configuration is generated through a small random
variation; then the system evolves to the new configuration with
probability 1 if its action is smaller, or else with probability
$\exp(-\Delta S/\hbar)$, and so on.

When the ``bare" parameters of the action are changed, the ground state
of the system, corresponding to the minimum of the action, changes too
(``phases"  of the theory). It is possible to insert first some bare
parameters into the action, then follow the evolution of the system towards
its ground state, and here measure the effective average value of the same
parameters.  The effective coupling constant, for instance, is usually
extracted in this way from the measured potential $U$. 

An interesting application of this method is the quantum Regge calculus by
Hamber and Williams \cite{hw}. In this case, the physical system under
investigation is very peculiar: the Euclidean spacetime, represented by a
simplicial manifold and numerically coded in terms of edge lengths and
defect angles (see also Section \ref{how} below). In order to obtain from
this discretized theory a continuum limit, independent of the details of
the discretization procedure, one looks in the parameters space of the
system for a second-order transition point, where the range of field
correlations diverges. 

Does this Euclidean model of the gravitational field constitute a faithful
representation of real spacetime, with its complex causal structure
distorted by field fluctuations? This question is still unanswered. While
we know for sure that for certain curved manifolds the analytical
continuation to Euclidean signature is not valid \cite{eqg}, there are no
theorems that do allow this continuation in some special case. All we can
do is hope that for weak fluctuations with respect to flat space, the Wick
rotation of real time to the imaginary axis still makes sense. 

Particle physicists do not doubt that the Euclidean Einstein action for
weak fields represents a massless spin 2 field correctly and in an unique
way. This point of view about gravitation, at variance with the
geometrodynamical view of spacetime, has been supported, as is well known,
by Feynman, Weinberg and others (see for instance the review by Alvarez
\cite{alv}). No problems have ever been encountered -- apart from the
familiar non-renormalizability of Einstein action -- in Euclidean
perturbation theory around flat space. The Euclidean formula for the
potential works well, too \cite{for}. 

Nevertheless, a serious problem still affects Euclidean quantum gravity,
even in the weak field sector:  the non positivity of the action. Either
if one takes the geometrical point of view (``the Euclidean action is not
at a minimum for $R=0$...") or the particle-physicist point of view (``the
quadratic part of the action has undefined sign..."), one ends up in the
unpleasant situation of studying perturbatively a system around a
configuration that does not appear to be a minimum for the action, but
rather a saddle point. The feeling is to control only one part of the
dynamics of the system, while the other part -- which makes the weak field
approximation work -- remains elusive.

The non-perturbative Euclidean quantum Regge calculus based upon Einstein
action can be helpful under this respect. It represents a geometrical
model whose dynamics is entirely under control, at least numerically. So
one can use it to throw some light on the paradoxes of the continuum
Einstein action. 

The short-distance regime of Euclidean quantum gravity is to some extent
arbitrary, and a strict connection with real spacetime is quite unlikely
in that limit. (We recall that due to the dimensional structure of the
gravitational action, field fluctuations are stronger at short distances.)
In the large distance limit, however, the connection is much more
plausible and thus it can be interesting to see how flat space emerges and
keeps stable in the Euclidean Regge calculus. 

The plan of the paper is the following.  In Section \ref{rec} we discuss
the general formulas which relate, in Euclidean field theory, the static
potential of two sources to the vacuum correlations of the field. This
also gives us the chance to introduce some basic notations.  In Section
\ref{dif} we deal with the non-positivity of the Euclidean action, and
give explicit examples in the weak field approximation.  In Section
\ref{zer} we present a novel issue: the zero-modes of the integral of the
scalar curvature.  In Section \ref{how} we give an interpretation of the
numerical results of Regge calculus in view of the stability problem. We
stress the importance of the sign of the effective cosmological term,
which acts as a volume term. In Section \ref{ads} we check the
geometrical argument of Section \ref{how} in the continuum, doing a
stability analysis of anti-de Sitter space.

\section{Static potential and Euclidean vacuum correlations.}
\label{rec}

In Euclidean quantum field theory there is a simple connection between the
static potential associated to a bosonic field and the vacuum correlations
of the field. This allows signs to be fixed without any ambiguity, which
is often crucial in stability issues. 

Consider a system comprising a quantum field and an external source $J$,
and denote by $W[J]=\langle 0^+|0^-\rangle_J$ the vacuum-to-vacuum
transition amplitude in the presence of the source. The energy of the
ground state of the system is given by
	\begin{equation}
	E_0 = -\frac{\hbar}{T} \log W[J],
\label{uno}
\end{equation}
	where $T$ is the temporal range of the source, which eventually
tends to infinity.  To check this, let us insert a complete set of energy
eigenstates $\{|n \rangle \}$ into the amplitude $W[J]$: 
	\begin{eqnarray}
	\langle 0^+|0^- \rangle_J & = &
	\langle 0| e^{-HT/\hbar} |0 \rangle \nonumber \\
	& = & \sum_n \langle 0| e^{-HT/\hbar} |n \rangle  \langle n|0 
	\rangle \nonumber \\ 
	& = & \sum_n | \langle 0|n \rangle |^2 e^{-E_nT/\hbar} .
\label{due}
\end{eqnarray}
	The smallest eigenvalue among the $E_n$'s corresponds to the
ground state, and in the limit $T \to \infty$ its exponential dominates
the sum. Thus taking the logarithm of $W[J]$ and multiplying by
$(-\hbar/T)$ we obtain the eigenvalue itself. One often considers
pointlike sources kept at rest at ${\bf x}_1...{\bf x}_N$, namely
	\begin{equation}
	J(x) = \sum_{j=1}^N q_j \, \delta^3 \left( {\bf x}-{\bf x}_j 
	\right).
\label{eq3}
\end{equation}
	In this case the ground state energy corresponds, up to a possible
additive constant, to the static potential $U$ of the interaction of the
sources, and depends on ${\bf x}_1...{\bf x}_N$. We have
	\begin{equation}
     U({\bf x}_1...{\bf x}_N) = -\frac{\hbar}{T} 
     \log \langle \exp(-S_J) \rangle_0 ,         
\label{a}
\end{equation}
	where $S_J$ is the term in the action containing the coupling to
$J$, and the average $\langle ... \rangle_0$ is computed through the
functional integral, weighing the field configurations with the factor
$\exp(-S_0/\hbar)$. For a scalar field $\phi$, $S_J$ takes the form
	\begin{equation}
     S_J = \int d^4x \, J(x) \phi(x) 
\end{equation}
	and $J(x)$ is as in (\ref{eq3}). For a gauge field $A_\mu$, the
source term is
	\begin{eqnarray}
     S_J & = & \int d^4x \, J_\mu(x) A_\mu(x); \\
     J_\mu(x) & = & \sum_{j=1}^N q_j \, \delta_{0 \mu} \,
     \delta^3 \left( {\bf x}-{\bf x}_j \right) .
\end{eqnarray}
	Therefore $S_J$ reduces to a sum of one-dimensional integrals
computed along temporal lines. When the pointlike sources are only two and
the field $A_\mu$ vanishes at infinity, $U({\bf x}_1,{\bf x}_2)$ can be
also expressed in terms of a Wilson loop. For gravity the source term is
	\begin{eqnarray}
	S_T & = & \frac{1}{2} \int d^4x \, 
	\sqrt{g(x)} \, T_{\mu \nu}(x) h_{\mu \nu}(x); \\ \label{lin}
	T_{\mu \nu}(x) & = & \sum_{j=1}^N m_j \, 
	\delta_{\mu 0} \, \delta_{\nu 0} 
	\, \delta^3({\bf x}-{\bf x}_j).
\end{eqnarray}
	The leading order contribution to (\ref{a}) in the case of two
pointlike sources is given in general by an expression of the form
	\begin{eqnarray}
     & & U({\bf x}_1,{\bf x}_2) = -\frac{\hbar}{T} 
     \, q_1 q_2 \nonumber \\
	& & \times \int_{-T/2}^{T/2} dt_1 \int_{-T/2}^{T/2} dt_2 
     \, \langle \Phi(t_1,{\bf x}_1)\Phi(t_2,{\bf x}_2) \rangle ,
\label{dop}
\end{eqnarray}
	where $\langle ... \rangle$ denotes the free propagator and $\Phi$
corresponds to the field $\phi$ in the scalar case and to the components
$A_0$ and $h_{00}$ in the electromagnetic and gravitational cases,
respectively (in the latter case, $q_1$ and $q_2$ are replaced by the
masses $m_1$ and $m_2$ of the sources). 

The sign of the correlation $\langle \Phi(t_1,{\bf x}_1)\Phi(t_2,{\bf
x}_2) \rangle$ is directly related to that of the potential energy. Some
care is needed in order to pick the correct convention for the Euclidean
metric. Eq.s (\ref{uno}), (\ref{due}) hold for the Euclidean metric with
signature (-1,-1,-1,-1), which is directly connected to the standard
Minkowski metric (1,-1,-1,-1) by the transformation $x_0 \leftrightarrow
ix_0$.  If the Euclidean metric (1,1,1,1) is used, eq.\ (\ref{dop}) holds
with the + sign.  This metric is usually preferred and will be employed in
the following. 

In the scalar and electromagnetic case, the correlation is positive. One
finds (apart from positive numerical factors and with $c\equiv 1$)
	\begin{equation}
	\langle \phi(x_1)\phi(x_2) \rangle  \sim 
	\langle A_0(x_1)A_0(x_2) \rangle  \sim 
	\frac{\hbar^{-1}}{(x_1-x_2)^2}; 
\label{corem}
\end{equation}
	thus the potential is repulsive if $q_1$ and $q_2$ have the same
sign, since 
	\begin{eqnarray}
	& & \int_{-T/2}^{T/2} dt_1 \int_{-T/2}^{T/2} dt_2 
	\frac{1}{(t_1-t_2)^2+({\bf x}_1
	- {\bf x}_2)^2} \nonumber \\
	& & \sim \frac{T}{|{\bf x}_1-{\bf x}_2|} .
\end{eqnarray}
	In the gravitational case, the correlation is negative:
	\begin{equation}
	\langle h_{00}(x_1)h_{00}(x_2) \rangle  \sim - 
	\frac{\hbar^{-1}G}{(x_1-x_2)^2}; 
\label{corgra}
\end{equation}
	being $m_1$ and $m_2$ always positive for physical sources, it
follows that the potential is always attractive. The negative sign of the
correlation (\ref{corgra}) may look counterintuitive. One should never
forget, however, that quantum fields are distributions, and the analogy
between a quantum functional integral and classical fields at finite
temperature has only a limited validity. 

The positivity of the scalar action and of the electromagnetic action in
Feynman gauge are evident in the Euclidean theory: one has namely in
momentum space 
	\begin{eqnarray}
	S_\phi & \sim & \int d^4p \, p^2 \, 
	\tilde{\phi}^*(p) \tilde{\phi}(p) , \\
	S_A & \sim & \int d^4p \, p^2 \, \delta_{\mu \nu} 
	\tilde{A}^*_\mu(p) \tilde{A}_\nu(p)
\end{eqnarray}
	and for the propagators
	\begin{equation}
	\tilde{G}_\phi(p) \sim p^{-2}; \ \ \ \ \ 
	\tilde{G}_{A,\mu \nu}(p) 
	\sim \delta_{\mu \nu} \, p^{-2} . 
\end{equation} 
	[We recall that, still apart from positive numerical factors,
$\int d^4p \, e^{ipx}p^{-2} \sim x^{-2}$. Compare (\ref{corem})].

\section{Different aspects of the same problem: the action does not have a
minimum.}
\label{dif}

The Hilbert-Einstein action for the gravitational field $g_{\mu \nu}(x)$
is usually written in the form
	\begin{equation}
     S = - \frac{1}{8\pi G} \int d^4x \, \sqrt{g(x)} R(x),
\label{act}
\end{equation}
where $R(x)$ is the scalar curvature.

Naively one can observe already at this stage that since $R$ is a quantity
which can be positive as well as negative and contains the first and
second derivatives of the metric, the integrand does not have a definite
sign and can grow in both directions if $g_{\mu \nu}(x)$ varies strongly. 

Hawking showed formally several years ago \cite{haw} that the Euclidean
action is not bounded from below \cite{nota}.
His argument is
important also because it does not make any reference to the weak field
approximation. Several possible solutions to the unboundedness problem
were proposed later on \cite{unb}. 

Wetterich suggested recently \cite{wet} a non-local modification of the
effective Euclidean action and showed that the phenomenological
implications of such a modified action are almost entirely compatible with
cosmology. Without entering into this matter, we just quote here his
decomposition of the tensor of the metric fluctuations $h_{\mu \nu}(x)
=g_{\mu \nu}(x)- \delta_{\mu \nu}(x)$ in terms of irreducible
representations of the Euclidean group in $d$ dimensions:
	\begin{eqnarray}
     h_{\mu \nu}(x) &=& b_{\mu \nu}(x) + \partial_\mu a_\nu(x) + 
     \partial_\nu a_\mu(x) \nonumber \\
     & & + \left( \partial_\mu \partial_\nu - 
     \frac{1}{d} \delta_{\mu \nu} \partial^2 \right) \chi(x)
     + \frac{1}{d} \delta_{\mu \nu} \sigma(x) ,
\end{eqnarray}
where the tensors $b_{\mu \nu}(x)$ and $a_\mu(x)$ satisfy the conditions
	\begin{equation}
     \partial_\mu a_\mu(x) = 0, \ \ \ \ \partial_\mu b_{\mu \nu}(x) = 0,
     \ \ \ \ \delta_{\mu \nu} b_{\mu \nu}(x) = 0 .
\end{equation}
To second order in $h_{\mu \nu}$ one obtains
	\begin{eqnarray}
     \sqrt{g(x)} R(x) &=& \frac{1}{4} \partial_\rho b_{\mu \nu}(x) 
     \partial_\rho b_{\mu \nu}(x)
     -\frac{(d-1)(d-2)}{4d^2} \times \nonumber \\
     & & \times \partial_\mu \left[ \sigma(x) - 
     \partial^2 \chi(x) \right] \partial_\mu \left[ \sigma(x)
     - \partial^2 \chi(x) \right] . \nonumber
\end{eqnarray}
	Therefore for $d>2$ the action becomes negative semi-definite for
configurations in which $b_{\mu \nu}(x)$ is zero. It can be shown that the
addition of a gauge-fixing term does not change the situation.  Wetterich
also observes that even though it is possible to make the action positive
definite adding short-distance terms (like the $R^2$ term), the effective
action, relevant for large distances, will always keep non positive.

In certain cases it can be reasonable to introduce in the theory a cut-off
on the momenta, and this will make the scalar curvature bounded. Still,
the quadratic part of the action will not be positive-definite. This
unpleasant feature does not only concern the small distances sector. It
can be exhibited most clearly in harmonic gauge. In this gauge the
quadratic part of the Hilbert-Einstein lagrangian in momentum space is 
simply given by
	\begin{equation}
     \tilde{L}^{(2)}(p) 
     \sim -p^2 \, \tilde{h}^*_{\mu \nu}(p) \, V_{\mu \nu \alpha \beta} 
     \, \tilde{h}_{\alpha \beta}(p) ,
\end{equation}
	where $V_{\mu \nu \alpha \beta}$ is a constant tensor which in
particular in dimension 4 is equal to
	\begin{equation}
     V_{\mu \nu \alpha \beta} = \delta_{\mu \alpha} \delta_{\nu \beta} +
     \delta_{\mu \beta} \delta_{\nu \alpha} - \delta_{\mu \nu} 
     \delta_{\alpha \beta}.
\end{equation}

Let us rearrange the 10 independent components of the tensor $\tilde{h}$, to
build an array $\tilde{h}_i$ ($i=0,1,...,9$), as follows: $\tilde{h}_{00} 
\to \tilde{h}_0,\ ...\ ,\tilde{h}_{33} \to \tilde{h}_3,\ \tilde{h}_{01} \to 
\tilde{h}_4,\ ...\ ,\tilde{h}_{23} \to \tilde{h}_9$. We then have
	\begin{equation}
     \tilde{L}^{(2)}(p) 
     \sim -p^2 \, \tilde{h}^*_i(p) \, M_{ij}
     \, \tilde{h}_j(p) ,
\end{equation}
where {\bf M} is a block matrix of the form
	\begin{equation}
     {\bf M} = \left[ \begin{array}{cc}
     {\bf m}(4 \times 4) & {\bf 0}(4 \times 6) \\
     {\bf 0}(6 \times 4) & {\bf 1}(6 \times 6) \end{array} \right] 
\end{equation}
and
	\begin{equation}
     {\bf m} = \left[ \begin{array}{rrrr}
     1 & -1 & -1 & -1 \\
     -1 & 1 & -1 & -1 \\
     -1 & -1 & 1 & -1 \\
     -1 & -1 & -1 & 1 \end{array} \right] .
\end{equation}
One easily checks that ${\bf m}^2=4 \times {\bf 1}$, thus the propagator of 
$\tilde{h}_i$ is given by
\begin{equation}
	\tilde{G}_{h}(p) \sim - p^{-2} \, {\bf M}^{-1} =
     - p^{-2} \, \left[ \begin{array}{cc}
     \frac{1}{4}{\bf m} & {\bf 0} \\
     {\bf 0} & {\bf 1} \end{array} \right] .
\end{equation}
	As anticipated in Section \ref{rec}, we see here that the
correlation function of $h_{00}$, like the other ``diagonal" correlations,
is negative. 

In order to check that the quadratic part of the gravitational action is 
not positive-definite, we can also rewrite $\tilde{L}^{(2)}(p)$ in matrix 
form as
	\begin{equation}
     \tilde{L}^{(2)}(p) \sim -p^2 \left\{ 2 {\rm Tr} \, [\tilde{h}^*(p) 
     \tilde{h}(p)] - |{\rm Tr} \, \tilde{h}(p)|^2 \right\}.
\end{equation}
Denoting now by $\tilde{h}_A(p)$ $(A=0,1,2,3)$ the eigenvalues of the 
symmetric matrix $[\tilde{h}_{\mu \nu}(p)]$, we have
	\begin{equation}
     \tilde{L}^{(2)}(p) \sim -p^2
     \left[ \sum_A |\tilde{h}_A(p)|^2 - \sum_{A \neq B} 
     \tilde{h}^*_A(p) \tilde{h}_B(p) \right] ,
\end{equation}
or
	\begin{equation}
     \tilde{L}^{(2)}(p) \sim -p^2 \, 
     \tilde{h}^*_A(p) \, m_{AB} \, \tilde{h}_B(p).
\label{quad}
\end{equation}
The eigenvalues of ${\bf m}$ are found to be (2,2,-2,-2). Thus the quadratic 
form (\ref{quad})  has no definite sign.

\section{The ``zero modes" in the integral of $R$.}
\label{zer}

It is known that if the metric has Lorentzian signature, then Einstein
equations in vacuum admit wave-like solutions. The Riemann tensor
$R^\mu_{\nu \rho \sigma}$ propagates in these solutions, while the Ricci
tensor $R_{\mu \nu}$ and the curvature scalar $R$ are identically zero. 

We recall that the Einstein equations in the presence of a source $T_{\mu
\nu}$ are (with $c \equiv 1$)
	\begin{equation}
	R_{\mu \nu}(x)- \frac{1}{2} g_{\mu \nu}(x) R(x) 
	= -8 \pi G T_{\mu \nu}(x),
\label{ein}
\end{equation}
and their trace is
	\begin{equation}
	R(x)=8 \pi G {\rm Tr} \, T(x).     
\label{tra}
\end{equation}
	From (\ref{tra}) we see that if $T_{\mu \nu}=0$, then $R=0$, as 
mentioned; therefore gravitational waves are a set of ``zero modes" of the
Hilbert-Einstein {\it lagrangian} \cite{pass}.

There is, however, another peculiar way to obtain zero modes of the
gravitational {\it action}. This is due to the non-positivity of this
action. 

Let us consider a solution $g_{\mu \nu}$ of eq.\ (\ref{ein}) with a
(covariantly conserved) source $T_{\mu \nu}$ obeying the additional
integral condition
	\begin{equation}
	\int d^4x \, \sqrt{g(x)} \, {\rm Tr} \, T(x) = 0.  
\label{add}
\end{equation}
	Taking into account eq.\ (\ref{tra}) we see that the action
(\ref{act})  computed for this solution is zero.  Condition (\ref{add})
can be satisfied by energy-momentum tensors that are not identically zero,
provided they have a balance of negative and positive signs, such that
their total integral is zero. Of course, they do not represent any
acceptable physical source, but the corresponding solutions of (\ref{ein}) 
exist nonetheless, and are zero modes of the action.

As an example of an unphysical source which satisfies (\ref{add}) one can
consider the static field produced by a ``mass dipole". Certainly negative
masses do not exist in nature; here we are interested just in the formal
solution of (\ref{ein}) with a suitable $T_{\mu \nu}$, because for this
solution we have $\int d^4x \, \sqrt{g} R=0$.  Let us take the following
$T_{\mu \nu}$ of a static dipole centered at the origin ($m,m'>0$): 
	\begin{equation}
	T_{\mu \nu}({\bf x}) = \delta_{\mu 0} \, \delta_{\nu 0} 
	\left[ m f( {\bf x}+{\bf a} ) 
	- m' f( {\bf x}-{\bf a} ) \right] .
\label{dip}
\end{equation}
	Here $f({\bf x})$ is a smooth test function centered at ${\bf x}=0$,
rapidly decreasing and normalized to 1, which represents the mass density.
The range of $f$, say $r_0$, is such that $a \gg r_0 \gg r_{Schw}$, where 
$r_{Schw}$ is the Schwartzschild radius corresponding to the mass $m$.
The mass $m'$ is in general different from $m$ and chosen in such a way
to compensate a possible small difference, due to the $\sqrt{g}$ factor,
between the integrals
	\begin{eqnarray}
	I^+ &=& \int d^4x \, \sqrt{g(x)} 
	f( {\bf x}+{\bf a} ) \ \ \  
	{\rm and} \nonumber \\  
	I^- &=& \int d^4x \sqrt{g(x)} 
	f( {\bf x}-{\bf a} ).
\label{piu}
\end{eqnarray}

The procedure for the construction of the zero mode corresponding to the
dipole is the following. One first considers Einstein equations with the
source (\ref{dip}). Then one solves them with a suitable method, for
instance in the weak field approximation. Finally, knowing $\sqrt{g(x)}$
one computes the two integrals (\ref{piu}) and adjusts the parameter $m'$
in such a way that $(mI^+ - m'I^-)=0$. 

Let us implement this procedure to first order. Inside a single mass 
distribution $mf({\bf x})$, with radius $r_0$ such that $r_0 \gg r_{Schw}$,
the gravitational field satisfies a static equation whose linear 
approximation is of the form
	\begin{equation}
	D h({\bf x}) = m \kappa f({\bf x}),
\label{linear}
\end{equation}
	where $D$ is a linear partial differential operator and $\kappa$
denotes, for brevity, $8\pi G$. Let us call $\hat{h}({\bf x})$ the
solution of (\ref{linear}) with $m\kappa$ replaced by 1. The solution of
the linearized Einstein equations with the source (\ref{dip}) is, in the
region with positive density, $h^+({\bf x})= m\kappa\hat{h}({\bf x}+{\bf
a})$. In the region with negative density the solution is $h^-({\bf x})=
-m'\kappa\hat{h}({\bf x}-{\bf a}/2)$. Thus in the region with positive
density we have
	\begin{equation}
	\sqrt{g(x)} \sim 1 + \frac{1}{2} m \kappa \, {\rm Tr} \, 
	\hat{h} ( {\bf x}+{\bf a} )
\end{equation}
and in the region with negative density
\begin{equation}
	\sqrt{g(x)} \sim 1 - \frac{1}{2} m' \kappa \, {\rm Tr} \, 
	\hat{h} ( {\bf x}-{\bf a} ) .
\end{equation}

The value of the action functional corresponding to this linearized 
``virtual dipole" metric is 
\begin{eqnarray}
	& & - \frac{1}{\kappa} \int d^4x \, \sqrt{g(x)} R(x) =
	- \int d^4x \, \sqrt{g(x)} \, {\rm Tr} \, T(x) = \nonumber \\
	& & = - \int d^4x \, \sqrt{g(x)} \left[ m 
	f( {\bf x}+{\bf a} ) - m' 
	f( {\bf x}-{\bf a} ) \right] = \nonumber \\
	& & = - \int d^4x \, \left\{ \left[ 1 + 
	\frac{1}{2} m \kappa \, {\rm Tr} \, 
	\hat{h} ( {\bf x}+{\bf a} ) \right]
	m f( {\bf x}+{\bf a} ) + \right. \nonumber \\
	& & \ \ \ \ - \left. \left[ 1 - \frac{1}{2} m' \kappa \, 
	{\rm Tr} \, \hat{h} ( {\bf x}-{\bf a} ) \right]
	m' f( {\bf x}-{\bf a} ) \right\} = \nonumber \\
	& & = - \int d^4x \, \left[ m f( {\bf x}+{\bf a} ) -
	m' f( {\bf x}-{\bf a} ) \right] + \nonumber \\
	& & \ \ \ \ - \frac{1}{2} \kappa \int d^4x \, \left[ m^2
	\, {\rm Tr} \, \hat{h} ( {\bf x}+{\bf a} )
	f( {\bf x}+{\bf a} ) + \right. \nonumber \\
	& & \ \ \ \ + \left. {m'}^2
	\, {\rm Tr} \, \hat{h} ( {\bf x}-{\bf a} )
	f( {\bf x}-{\bf a} ) \right].
\label{lunga}
\end{eqnarray}
	Being $f$ normalized, the first integral of (\ref{lunga}) gives 
$-T(m-m')$, where $T$ is the temporal integration range. We then have
	\begin{eqnarray}
	& & - \frac{1}{\kappa} \int d^4x \, \sqrt{g(x)} R(x) =
	- T(m-m') \nonumber \\
	& & \ \ \ \ - \frac{1}{2} \kappa T (m^2+{m'}^2)
	\int d^3x \, \, {\rm Tr} \, \hat{h}({\bf x}) f({\bf x}).
\end{eqnarray}
	The integral on the r.h.s.\ is a number of the order of 1 and 
will be denoted by $\eta$. The condition for a zero mode now reads
	\begin{equation}
	(m-m') + \frac{1}{2} \eta \kappa (m^2+{m'}^2) = 0
\end{equation}
and it is satisfied, up to terms of order $\kappa^2$, for $m'=m(1+\eta 
\kappa m)$.

\medskip
There is no obstacle, in the functional integral, to the formation of a
zero mode like this. It can ``pop up" at any point in spacetime, or more
likely it can be induced by an external localized source, even if weak.
The spatial size of the mode can be in principle arbitrarily large. 

These modes can develop both in Lorentzian and in Euclidean metric. 
Sometimes it is argued that in the functional integral with real time and
with the oscillating factor $\exp(iS/\hbar)$ the non-positivity of the
action has no importance. But also in that case the zero modes described
above can be present.

The only mechanism able to suppress these modes appears to be the presence
of an effective volume term with $\Lambda<0$ (see Section \ref{how}).

\section{How flat space emerges from the quantum Regge lattice.}
\label{how}

It can be helpful to recall briefly here the main features of the quantum
Regge calculus technique by Hamber and Williams \cite{hw}. In this
approach the Euclidean 4D spacetime is approximated by a simplicial
manifold and the curvature, all concentrated at the ``hinges", is
proportional to the defect angle which one finds when a hinge is flattened
out. The system is numerically simulated, with the edge lengths as
fundamental variables. At the beginning one puts into the action, as
``bare" parameters, $k$ (inverse of the Newton constant) and $a$
(coefficient of the $R^2$ term). Then one looks in the phase diagram of
the theory for a second-order transition point. 

It turns out that the phase diagram is divided in two regions: a ``smooth" 
phase, with average curvature small and negative, and fractal dimension
close to 4; and a ``rough" phase, singular, collapsed, with average
curvature large and positive and small fractal dimension. It is clear that
the sign of the curvature plays a crucial role in the stability of the
system. The two phases are separated by a transition line. Approaching
this line from the smooth phase, the average curvature $\langle R \rangle$
tends to zero. In this way, flat space is obtained in a dynamical way from
the smooth phase, without any need of introducing into the theory a flat
background by hand. From here, perturbation theory can somehow start.

A few data can help to complete the picture. The lattice sites are $16
\times 16 \times 16 \times 16=65,536$, with 1,572,864 simplices. The edge
lengths are updated by a straightforward Montecarlo algorithm. Eventually
an ensemble of configurations is generated, distributed according to the
Euclidean action. The topology is fixed as a four-torus with periodic
boundary conditions. A stable, well behaved ground state is found for
$k<k_c\sim 0.060$. The system resides in the smooth phase, with fractal
dimension four. Six values of $k$ have been investigated: 0.00, 0.01,
0.02, 0.03, 0.04, 0.05.

The static potential is attractive and can be fitted by $L^{-1}$, with a
small Yukawa factor $\exp(-mL)$. The mass extracted this way is consistent
with the exponential decay of the correlations of the scalar curvature. 
The effective Newton constant can be extimated to $G\sim 0.14$, in lattice
units. More exactly, at the beginning one puts $\lambda=1$ in the action.
Then one finds for the average edge length
	\begin{equation}
	l_0 = \sqrt{\langle l^2 \rangle} = 2.36, \ \ \ \ {\rm i.e.} 
	\ \ \ \ l_0 = 2.36 \, \lambda^{-1/4}. 
\label{latt}
\end{equation}
The critical value of the bare coupling is
	\begin{equation}
	k_c \sim 0.060, \ \ \ {\rm i.e.} \ \ \  k_c ~ 0.060 \, \lambda^{1/2}
\end{equation}
and the product $G k_c$ is independent of $l_0$ and finite, as hoped. The 
precision of these data is expected to improve considerably in the next 
months, thanks to the new dedicated supercomputer AENEAS \cite{aen}.

As far as stability is concerned, the numerical simulations show as
mentioned that in the smooth phase the system evolves toward a stable minimum
position with $\langle R \rangle <0$.  It is not hard to understand
intuitively, from the geometrical point of view, {\it why} the system is
stable in this phase.  

The effective action is
	\begin{equation}
     S_{eff} = \int d^4x \, \sqrt{g(x)} \left[ \frac{\Lambda}{8\pi G} 
     - \frac{R(x)}{8\pi G} \right] ,
\label{effa}
\end{equation}
	with $\Lambda \sim \langle R \rangle$. Let us assume that the
system is in a configuration with $R(x)=const. =\Lambda$. Now suppose that
somewhere a positive fluctuation of $R$ appears. Being proportional to
$\exp(-S_{eff})$, the probability of this new configuration is seemingly
larger, if we take into account only the second term of the action. The
first term, however (the effective cosmological term) can be written as
$(\Lambda {\cal V})$, where ${\cal V}$ is the total volume of the system. 
This volume is maximum when the manifold is flat and all hinges are
completely extended.  As soon as a curvature fluctuation appears at some
point the total volume decreases, and since $\Lambda$ is negative, this
tends to suppress the fluctuation. The converse happens, of course, in the
collapsed phase, where $\Lambda>0$.

Furthermore, this same mechanism will suppress in an even stronger way the
zero-modes described in Section \ref{zer}, since these modes cause no
variation of the integral of $R$, but only decrease the volume. 

The continuum theory is recovered from the lattice only at the transition
line, where $\Lambda =0$. This refers, however, to an average computed
over all space. More generally, $\Lambda$ scales with the volume $v$ of
the averaged region, according to a power law of the form 
$|\Lambda| G \sim (v^{-1/4}l_0)^\gamma$.

\section{Is AdS space a stable minimum of the continuum action?}
\label{ads}

	As we have seen in the previous Section, the discretized Euclidean
action appears to be stabilized by a negative cosmological term. This acts
as a volume term and opposes the curvature fluctuations, which tend to
diminish the volume of the lattice.
	
	Is the geometrical argument offered independent of the Regge
lattice regularization? This question suggests a check in the continuum.
If eq.\ (\ref{effa}) really has a stable minimum, then that minimum must
be a solution of the Euclidean Einstein equations with a negative
cosmological constant.  Such a solution is known; this is Euclidean
anti-de Sitter (AdS) space.

	Therefore, we can do a stability analysis: is AdS space a stable
minimum of the action, with only positive modes in the weak-field
expansion in this background?  Or is it only a saddlepoint, like flat
space? 

	A stability analysis in AdS space is more intricate than in flat
space. We must expand the metric with respect to the appropriate background,
namely
	\begin{equation}
	g_{\mu \nu}(x) = g^{AdS}_{\mu \nu}(x) + h_{\mu \nu}(x)
\end{equation}
	where $g^{AdS}_{\mu \nu}(x)$ is the solution of the vacuum
Euclidean Einstein equations with a negative cosmological term and thus
represents a space with constant negative curvature. The form of
the metric $g^{AdS}_{\mu \nu}(x)$ depends on the coordinates chosen. 
	
	We can formally expand the action as
	\begin{equation}
	S[g] = S[g^{AdS}] + \frac{\delta S}{\delta g} [g^{AdS}] \times h 
	+ \frac{1}{2} \frac{\delta^2 S}{\delta g^2} [g^{AdS}] \times h^2 
	+ ...
\end{equation}
	The first derivative vanishes at $g^{AdS}$ and to check the
stability we must study the sign of the quadratic form $U= \frac{\delta^2
S}{\delta g^2} [g^{AdS}]$. Remembering that the first variation of $S$
gives the Einstein equations, we obtain
	\begin{equation}
	U^{\alpha \beta \mu \nu}(x) = \left[ \frac{\delta S}
	{\delta g_{\alpha \beta}} \left( R^{\mu \nu} - \frac{1}{2} 
	g^{\mu \nu} R - \Lambda g^{\mu \nu} \right)
	\right]_{g=g^{AdS}(x)}.
\end{equation}
	The functional derivative of the first two terms corresponds, up
to a gauge-fixing, to the usual quadratic form of pure gravity, while the
derivative of the cosmological term gives $-\Lambda g^{AdS,\mu \alpha}(x)
g^{AdS,\nu \beta}(x)$ \cite{nota2}. 
	
	The second variation of $S$ at the extremum must take into account
the dependence of $U$ on $x$: 
	\begin{equation}
	\delta^2 S = \frac{1}{2} \int d^4x \, \sqrt{g^{AdS}(x)} 
	h_{\alpha \beta}(x) U^{\alpha \beta \mu \nu}(x) h_{\mu \nu}(x).
\end{equation}
	
	Since the AdS space is homogeneous and we are most interested in
localized fluctuations, we could restrict our attention to functions
$h_{\mu \nu}(x)$ having support in a small region around the origin. In
suitable coordinates we will have $g^{AdS}_{\mu \nu}(x) \sim g^{AdS}_{\mu
\nu}(0)= \delta_{\mu \nu}$, but the gauge fixing term constitutes a
serious problem, because it must be consistent with the symmetries of the
background and with the fact that the fluctuations are localized (compare
also Ref.\ \cite{tw}, for the de Sitter case and related horizon and
infrared problems). 
	
	In conclusion, investigating stability along these lines appears
to be very hard.

	Another possible check concerns the conformal mode.  In this case,
a weak field expansion is not necessary. For any conformal transformation
of the metric of the form
	\begin{equation}
	g_{\mu \nu}  \to  g'_{\mu \nu}(x) = \Omega^2(x) g_{\mu \nu}(x)
\end{equation}
the curvature scalar transforms as 
\begin{equation}
	R(x)  \to  R'(x) = \Omega^{-2}(x)R(x) - 6 \Omega^{-3}(x) 
	\partial^2 \Omega(x)
\end{equation}
and the action with cosmological term transforms as (we omit the 
$x$-dependence)
\begin{equation}
	S   \to  S' = - \frac{1}{8\pi G} \int d^4x \, \sqrt{g} 
	(\Omega^2 R + 6 \partial_\mu \Omega \partial_\nu \Omega 
	g^{\mu \nu} - \Lambda \Omega^4).
\label{stab}
\end{equation}
	Note that $g_{\mu \nu}(x)$ does not need to be constant, and in
our case coincides with the AdS metric.

	We see from (\ref{stab}) that for $\Lambda<0$ the conformal mode
is {\it not} stabilized. On the contrary, it seems that conformal
fluctuations can increase the total volume of space and are therefore
enhanced.  This continuum result is in bold contrast with our intuition of
the behavior of the lattice. The contrast could be possibly explained as
follows.

\medskip

	(i) It turns out from the numerical simulations that after the
simplicial lattice has reached its ground state and has stabilized, the
average length $l_0$ of the links (the ``bones" of the triangulation) 
keeps constant and fluctuations are small. 

	This behavior could be due in part to the $R^2$ term or to the
volumes in phase space, rather than to the Einstein $R$ term;  it signals,
anyway, that an effective suppression of the conformal modes has occurred. 

\medskip

	(ii) If the lengths of the lattice links are approximately
constant, then any increase of the curvature implies an increase of defect
angles and thus a diminution of the total volume, as argued in the
previous Section. 

	This is easily visualized in two dimensions. Let us consider, for
instance, an orthogonal pyramid with a regular polygon as its basis (but
here we are only interested in the side surface of the pyramid---the 2D
volume). Suppose to keep constant the edges of the pyramid---the links. 
When the height of the pyramid goes to zero, the side surface is maximum
and the defect angle $\delta$, associated to the curvature, is zero. (The
defect angle is that obtained by ``opening" the pyramid, as if it was made
of paper.) The sharper the pyramid, the larger $R \sim \delta$ and the
smaller the side surface.  The variation of the side surface is of the
order of $\Delta {\cal S} \sim - (\Delta \sin \delta) l_0^2$.

	The 4D analogue is $\Delta {\cal V} \sim - (\Delta \sin \delta)
l_0^4$, while the contribution of the curvature to the Einstein action is
of the order of $\Lambda (\Delta \delta) l_0^2$.  Since in lattice units
we have (compare (\ref{latt})) $l_0 = \sqrt{\langle l^2 \rangle} >1$ and
$|\Lambda|>1$, the lattice prefers to keep the $\delta$'s close to zero. 

\medskip
	The author would like to thank all the staff of ECT* for the kind
hospitality during completion of this work.


\begin{references}
\bibitem{nota}
	In the $R+R^2$ theory the action is positive definite, provided
the coefficient $a$ of the $R^2$ term is such that $4a\lambda-k^2>0$
(compare Section \ref{how}). Remembering that the $R^2$ term also makes
the theory renormalizable, it is easy to understand why the $R+R^2$ action
has been chosen, instead of the simple Einstein action, for non
perturbative investigations through the Regge discretization. (The
unitarity problem does not show up in the non perturbative approach.) It
was found, however, that the emergence in the quantum Regge calculus of a
stable ground state with zero average curvature does not strictly require
the $R^2$ term.
\bibitem{pass}
	We note in passing that if the signature of the metric is
Euclidean, then the equation $\partial^2 f=0$ does not really admit
wave-like solutions.  Instead, it has the form of a 4D Poisson equation
and its solutions are harmonic functions, which reduce to costants if
there are no sources.
\bibitem{nota2}	
	Setting $g^{AdS}_{\mu \nu}(x)=const.=\delta_{\mu \nu}$ we would
obtain the known ``naive" statement that a cosmological term in weak field
approximation amounts to a mass term for the graviton (see for instance
\cite{vel}). It is clear, however, that this statement is not rigorously
true, since in AdS space the metric $g^{AdS}$ which minimizes the action
is not flat.
\end{references}
\end{document}